\documentclass[a4paper,11pt]{article}
\usepackage{pos}
\usepackage{amsmath,graphicx,multirow,tabularx,physics}
\usepackage{davitex,slashed,listings}

\definecolor{airforceblue}{rgb}{0.36,0.54,0.66}
\definecolor{burgundy}{rgb}{0.5,0.0,0.13}
\definecolor{blue-violet}{rgb}{0.54,0.17,0.89}
\newcommand{\mcrit}{m_{\text{crit}}}
\newcommand{\muRW}{\mu_{\text{RW}}}
\newcommand{\TRW}{T_{\text{RW}}}
\newcommand{\chidisc}{\chi^{\text{disc}}}

\title{Lattice QCD at Imaginary Chemical Potential in the Chiral Limit}
\ShortTitle{Chiral QCD at Imaginary Chemical Potential}

\author[a]{D.~A. Clarke}
\author*[a]{Jishnu Goswami}
\author[a]{F. Karsch}
\author[a]{Anirban Lahiri}
\author[a]{M. Neumann}
\author[a]{C.~Schmidt}

\affiliation[a]{Fakult\"at f\"ur Physik, Universit\"at Bielefeld,\\
Bielefeld, Germany}

\emailAdd{dclarke@physik.uni-bielefeld.de}
\emailAdd{jishnu@physik.uni-bielefeld.de}
\emailAdd{karsch@physik.uni-bielefeld.de}
\emailAdd{alahiri@physik.uni-bielefeld.de}
\emailAdd{neumann@physik.uni-bielefeld.de}
\emailAdd{schmidt@physik.uni-bielefeld.de}

\abstract{We report on an ongoing study on the interplay between Roberge-Weiss (RW) 
and chiral transitions in simulations of (2+1)-flavor QCD with an imaginary chemical 
potential. We established that the RW endpoint belongs to the 3-$d$, $\Z_2$ universality 
class when calculations are done with the Highly Improved Staggered Quark (HISQ) action 
in the RW plane with physical quark masses. We also have explored a range of quark 
masses corresponding to pion mass values, $m_\pi\geq40$~MeV and found that the 
transition is consistent with $\Z_2$ universality class. We argue that observables 
that were usually used to determine the chiral phase transition temperature, e.g. 
the chiral condensate and chiral susceptibility, are sensitive to the RW transition 
and are energy-like observables for the $\Z_2$ transition, contrary to the 
magnetic-like (order parameter) behavior at vanishing chemical potential. Moreover 
the calculations performed at $m_\pi\sim40$~MeV also put a stringent constraint  for a 
critical pion mass at zero chemical potential for a possible first-order chiral 
phase transition.}

\FullConference{%
 The 38th International Symposium on Lattice Field Theory, LATTICE2021
  26th-30th July, 2021
  Zoom/Gather@Massachusetts Institute of Technology
}

\begin{document}
\maketitle

\section{Introduction}\label{sec:intro}

% The x-axis labels here need to be fixed
\begin{figure}[b]
\centering
\includegraphics[width=0.45\textwidth]{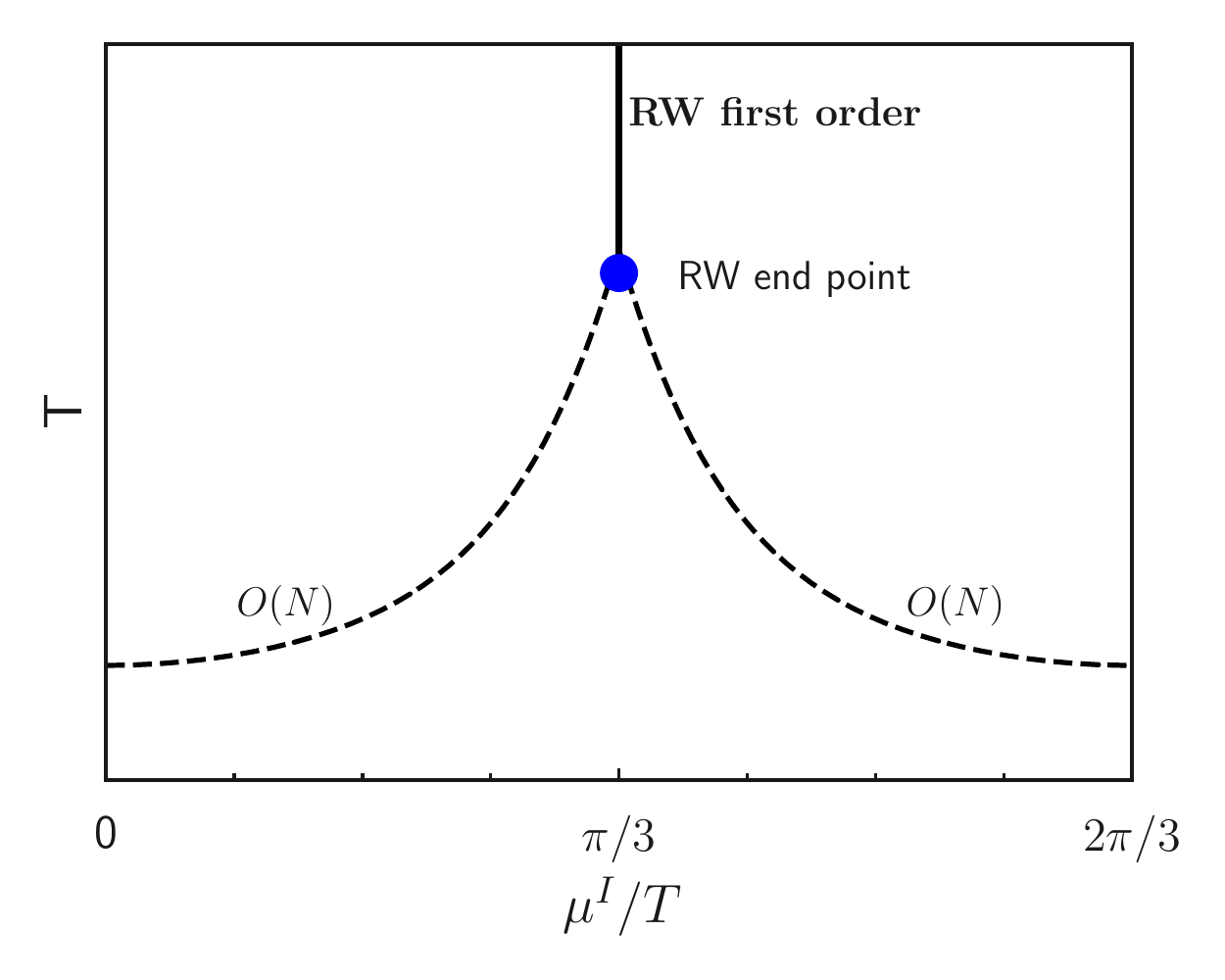}
\includegraphics[width=0.45\textwidth]{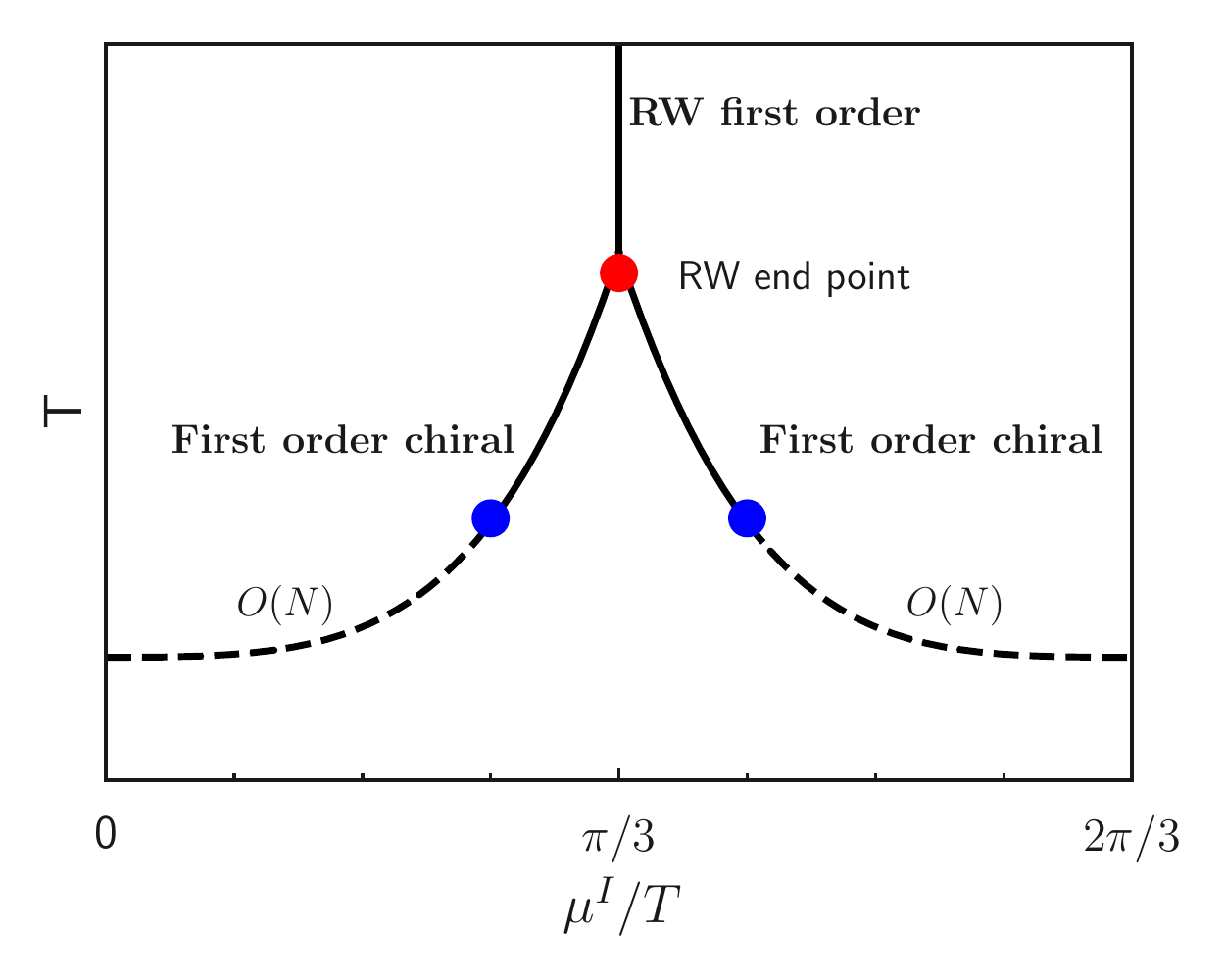}
\caption{Two possible phase diagrams in the $T$-$\mu_I$ plane at $m_l=0$. 
%The $x$-axis represents the imaginary part of the $\mu_I$. 
In both cases the vertical
line at high temperature is a first-order line at $\muRW$.
{\it Left}: An $\O(N)$ line 
emerges from $\mu_I=0$, terminating at 
a $\Z_2$ point, indicated in blue. 
{\it Right}: An $\O(N)$ line 
emerges from $\mu_I=0$, terminating again at a $\Z_2$ point.
This time, the transition continues as a first-order line, until
terminating at a first-order triple point, indicated in red.}
\label{fig:RWphase}
\end{figure}

We are exploring the phase diagram of QCD with two light, degenerate
flavors $l\equiv u=d$ and one heavier flavor $s$, {\it i.e.} $(2+1)$-flavor QCD. This phase diagram depends on
the temperature $T$, chemical potentials $\mu_f$ of the various quark
flavors $f$, and their masses $m_f$. 
At $m_l=0$ there is a chiral phase transition, and while it was originally
argued to be second-order belonging to the 3-$d$, $\O(4)$ universality 
class~\cite{pisarski_remarks_1984},
which is supported by recent lattice
calculations~\cite{HotQCD:2019xnw,Clarke:2020clx,Clarke:2020htu}, 
it is also possible that this
transition is first-order; indeed finally settling the nature of 
the chiral transition is still an open issue.

% Maybe it is also worth to mention the recent 3-flavor results here?

At $\mu_f=0$ and physical strange quark mass $m_s$, in the second-order
scenario, the transition is $\O(4)$
only at $m_l=0$ and crossover elsewhere. 
By contrast in the first-order scenario, an $\mcrit$ exists such that 
at $m_l=\mcrit$ the transition is $\Z_2$, with a first-order region
for $m_l<\mcrit$. In the case of three degenerate light quark flavors some evidence of this first-order region was found
from coarse lattices using an unimproved staggered discretization 
scheme~\cite{karsch_chiral_2001}; however this finding depends
strongly on the cutoff~\cite{Cuteri:2021ikv} and seems to disappear
under more highly improved 
discretizations~\cite{Bazavov:2017xul}. This also strongly suggests that the chiral phase transition in $(2+1)$-flavor QCD is second order.

The QCD partition function with a purely imaginary chemical
potential $\mu=i\mu_I$ is known to exhibit a $\Z_3$ periodicity~\cite{roberge_gauge_1986}
\begin{equation} 
  \mu_I/T\to\mu_I/T+2\pi n/3,
\end{equation}
where $n\in\Z$. Choosing $\mu_I$ at the center of this sector, i.e.
at $\mu_I/T=(2n+1)\pi/3$, is the Roberge-Weiss (RW) plane, and we denote the
corresponding chemical potential $\muRW$. 
In studies where one finds a first-order region, it is also found
that $\mcrit=\mcrit(\mu_I)$ increases with increasing, purely imaginary chemical
potential~\cite{Bonati:2014kpa}. This $\mcrit$ is largest at
$\mcrit(\muRW)$; therefore in the context of looking out for the
previously mentioned first-order region, it is useful to look out
for this $\mcrit$ in the RW plane, which can then be used to place an
upper bound on $\mcrit(0)$.

Two possible phase
diagrams in the $T$-$\mu_I$ plane are shown schematically in
Fig.~\ref{fig:RWphase}. In the second-order scenario, shown on the left,
a transition line, corresponding to pseudo-critical behavior for any non-zero value of the light quark masses and a phase transition in the $\O(N)$ universality class for vanishing light quark masses, starts at $\mu_I=0$ and terminates
at a $\Z_2$ end point on the RW plane. By contrast in the first-order
scenario, the RW endpoint must be first-order triple; a possible
way this could happen is shown on the right. In this case the line of first order phase transitions emerging from the triple-point corresponds to genuine first order chiral phase transitions and with decreasing quark mass values this region in parameter space could extend all the way down to $\mu_I=0$.

These proceedings give the current status of our ongoing
work~\cite{Goswami:2018qhc,goswami_critical_2019} investigating these aspects of
the chiral transition from the perspective of the RW plane.

\section{Renormalization group setup}\label{sec:RG}

Roberge and Weiss argued that the QCD partition function at imaginary chemical
potential is symmetric in $\mu_I$ about $\muRW$~\cite{roberge_gauge_1986}. 
This corresponds to a $\Z_2$ symmetry
that may spontaneously break above a critical temperature $\TRW$. We define the physical lattice volume $V=(N_\sigma^3 a)$ and the temperature $T=1/(N_\tau a)$, where $a$ is the lattice spacing.
The Polyakov loop on an $N_\sigma^3\times N_\tau$ lattice is given by
\begin{equation}
 P=\frac{1}{3N_\sigma^3}\sum_{\vec{x}}\tr\prod_\tau U_4(\vec{x},\tau) \; .
\end{equation}
The imaginary part of $P$ changes sign under $U\to U^\dagger$, while the QCD
action remains unchanged; hence $\ev{\Im P}$ can be used as an order parameter
for the RW transition at $T=\TRW$ and $\mu_I=\muRW$. In an effective Hamiltonian
written near this critical endpoint, $\Im P$ couples to the symmetry-breaking field
$h\equiv\mu_I-\muRW$. Observables that respect the symmetry will couple
to the reduced temperature $t\equiv(T-\TRW)/\TRW$. 

If this RW endpoint belongs to the 3-$d$, $\Z_2$ universality class, then in a
neighborhood of this point,
the logarithm of the partition function can be expressed as
\begin{equation}\label{eq:scalingfunc}
  f\sim
   b^{-3}f_{\text{s}}(b^{1/\nu}t/t_0,b^{\beta\delta/\nu}h/h_0,b^{-1}N_\sigma/l_0)
   + \text{reg.}
\end{equation}
The first term indicates the singular contribution, written in terms of the
scale factor $b$, universal critical exponents $\beta$, $\delta$, and $\nu$,
and non-universal scale parameters $t_0$, $h_0$, and $l_0$ as well as $T_{RW}$. The second term
indicates regular contributions, which can be written as a Taylor series in $t$,
$h$, and $N_\sigma$.

In this study, we examine (2+1)-flavor QCD on the RW plane, i.e. for $h=0$. 
Since $\ev{\Im P}$ vanishes at $h=0$ for all
$T$ in a finite volume, one may take as order parameter and corresponding susceptibility
% Other paper missing a color factor?
\begin{equation}
 M\equiv\ev{|\Im P|}~~~~~~~~~\text{and}~~~~~~~~~
  \chi_M\equiv N_\sigma^3\left(\ev{|\Im P|^2}-\ev{|\Im P|}^2\right).
\end{equation}
Furthermore since $h=0$, if we set $b=N_\sigma/l_0$, eq.~\eqref{eq:scalingfunc}
simplifies, and one can derive the scaling behavior
\begin{equation}\begin{aligned}\label{eq:Mscaling}
 M(T,V)&= AN_\sigma^{-\beta/\nu}f_{G,L}(z_f)+\text{reg.},\\
 \chi_M(T,V)&= A^2N_\sigma^{\gamma/\nu}f_{\chi,L}(z_f)+\text{reg.},
\end{aligned}\end{equation}
where $f_{G,L}$ and $f_{\chi,L}$ are finite size scaling functions for the order parameter
and susceptibility~\cite{Engels:2002fi} that depend on the finite size
scaling variable\footnote{We choose $l_0=1$, so $z_0=1/t_0$.} 
$z_f=z_0N_\sigma^{1/\nu}$ and $\gamma$ is another critical exponent.
In addition, we calculate the Binder cumulant~\cite{binder_finite_1981}
\begin{equation}\label{eq:Bscaling}
  B_4\equiv\frac{\ev{\Im P^4}}{\ev{\Im P^2}^2},
\end{equation}
which, near the critical point, is just a ratio $f_B$ of scaling functions,
\begin{equation}
  B_4(T,V)= f_B(z_f)+\text{reg.}
\end{equation} 

In the chiral limit, a chiral phase transition occurs for all $\mu_I$. To probe
this transition, one can use the renormalization-group-invariant order parameter,
\begin{equation}\label{eq:ccRGI}
  \Delta_{ls}=\frac{2}{f_K^4}\left(m_s\ev{\bar{\psi}\psi}_l-m_l\ev{\bar{\psi}\psi}_s\right),
\end{equation}
where $f_K$ is the kaon decay constant, the chiral condensate for quark flavor
$f$ is
\begin{equation}\label{eq:cc}
  \ev{\bar{\psi}\psi}_f=\frac{1}{4N_\sigma^3N_\tau}\ev{\tr M^{-1}_f},
\end{equation}
and $M_f$ is the corresponding staggered fermion matrix. A chiral
pseudocritical temperature $T_{pc}$ at nonzero $m_l$ is defined by
the peak as a function of $T$ in the disconnected part of 
the chiral susceptibility 
\begin{equation}\label{eq:discsus}
  \chi^{\text{disc}}=\frac{m_s^2}{4N_\sigma^3 N_\tau f_K^4}
                       \left(\ev{(\tr M_l^{-1})^2}-\ev{\tr M_l^{-1}}^2\right).
\end{equation}
This is only a part of the total chiral susceptibility; nevertheless peaks as a
function of temperature for $\chidisc$ and the total susceptibility will
coincide in the chiral limit.

In a neighborhood of $\TRW$, the observables in eq.~\eqref{eq:ccRGI} 
and \eqref{eq:discsus} will be affected by the RW transition. The
chiral condensate and susceptibility are even under $U\to U^\dagger$
and are thus expected to scale as an energy density and specific heat,
respectively. Similarly $m_l>0$
does not break the $\Z_2$ symmetry corresponding to the RW transition,
and is therefore an energy-like coupling. In particular this leads
to the expectation that $\chidisc$ will diverge in the infinite volume
limit as
\begin{equation}\label{eq:RGchidisc}
  \chidisc(T,V)\sim N_\sigma^{\alpha/\nu}f''_{f,L}(z_f),
\end{equation}
where $\alpha=2-d\nu$ is another 3-$d$, $\Z_2$ critical exponent and primes
indicate derivatives w. r. t. $z_f$.

%Finally, in addition to the susceptibilities $\chi_M$ and $\chidisc$
%of magnetization-like and energy-like order parameters, we will consider
%the mixed susceptibility
%\begin{equation}
%  \chi_t=\frac{m_s}{2f_K^4}\left(\ev{|\Im P|\tr M_l^{-1}}
%                                  -\ev{|\Im P|}\ev{\tr M_l^{-1}}\right)
%\end{equation}
%which is expected to scale as
%\begin{equation}
%  \chi_t(T,V)=A_t N_\sigma^{(1-\beta)/\nu}f_{G,L}'(z)+\text{reg.}
%\end{equation}

\section{Computational setup}\label{sec:setup}

% What are the statistics?
We perform our calculations using $(2+1)$-flavor QCD with HISQ fermions and the tree-level
improved Symanzik gauge action. We keep $m_s$ fixed to
its physical value and vary $m_l$ between $m_l=m_s/27$ to $m_s/320$,
which corresponds to a Goldstone pion mass between 135 and 40~MeV.
All lattices have $N_\tau=4$ and a finite imaginary chemical potential
$\mu/T=\pi/3$ for all quark flavors, i.e. we work on the first
RW-plane. A summary of simulation parameters
is given in Table~1 of Ref.~\cite{goswami_critical_2019}. We set the
scale at finite lattice spacing using the parameterization of the line of constant physics given 
in Ref.~\cite{Bazavov:2011nk,Bollweg:2021vqf}.

\section{Results}\label{sec:results}

\begin{figure}
\centering
\includegraphics[width=0.32\textwidth]{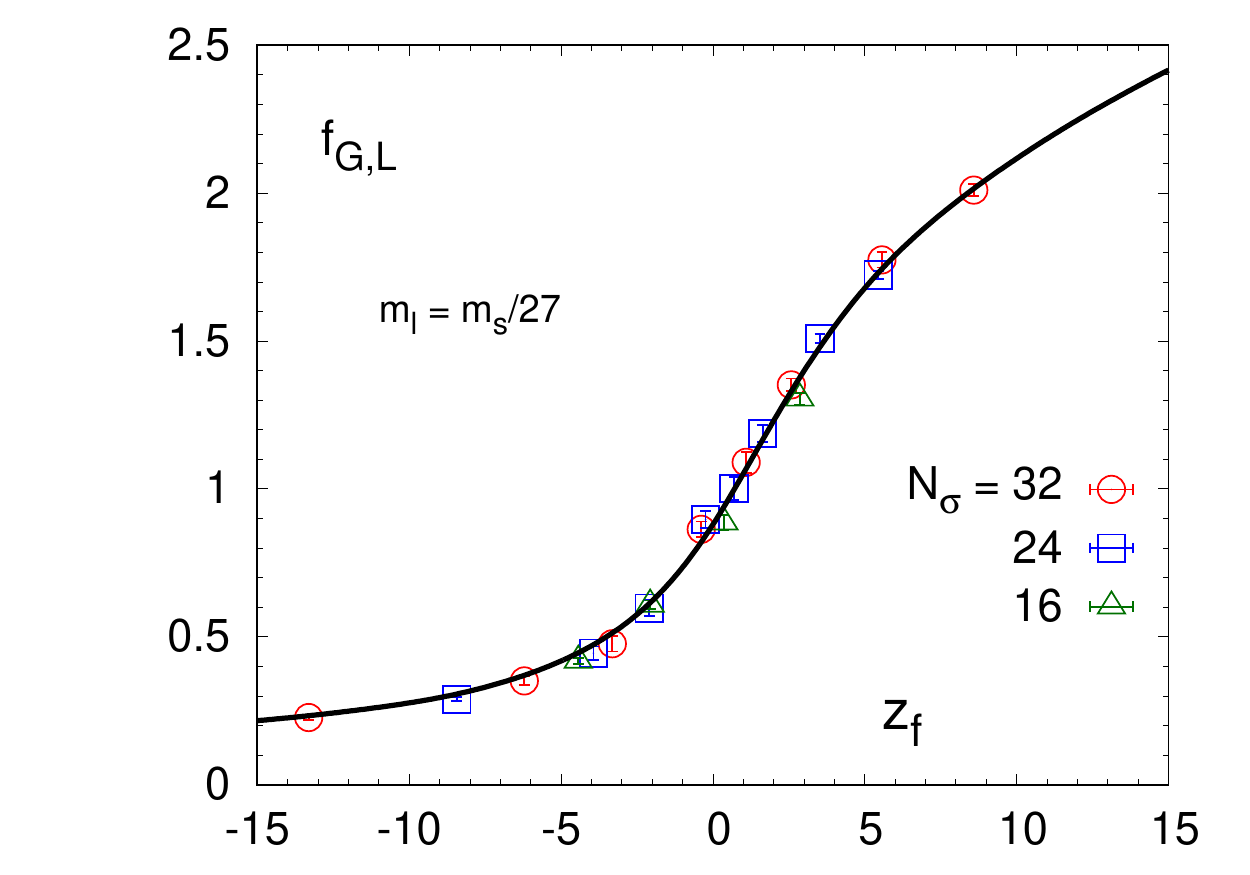}
\includegraphics[width=0.32\textwidth]{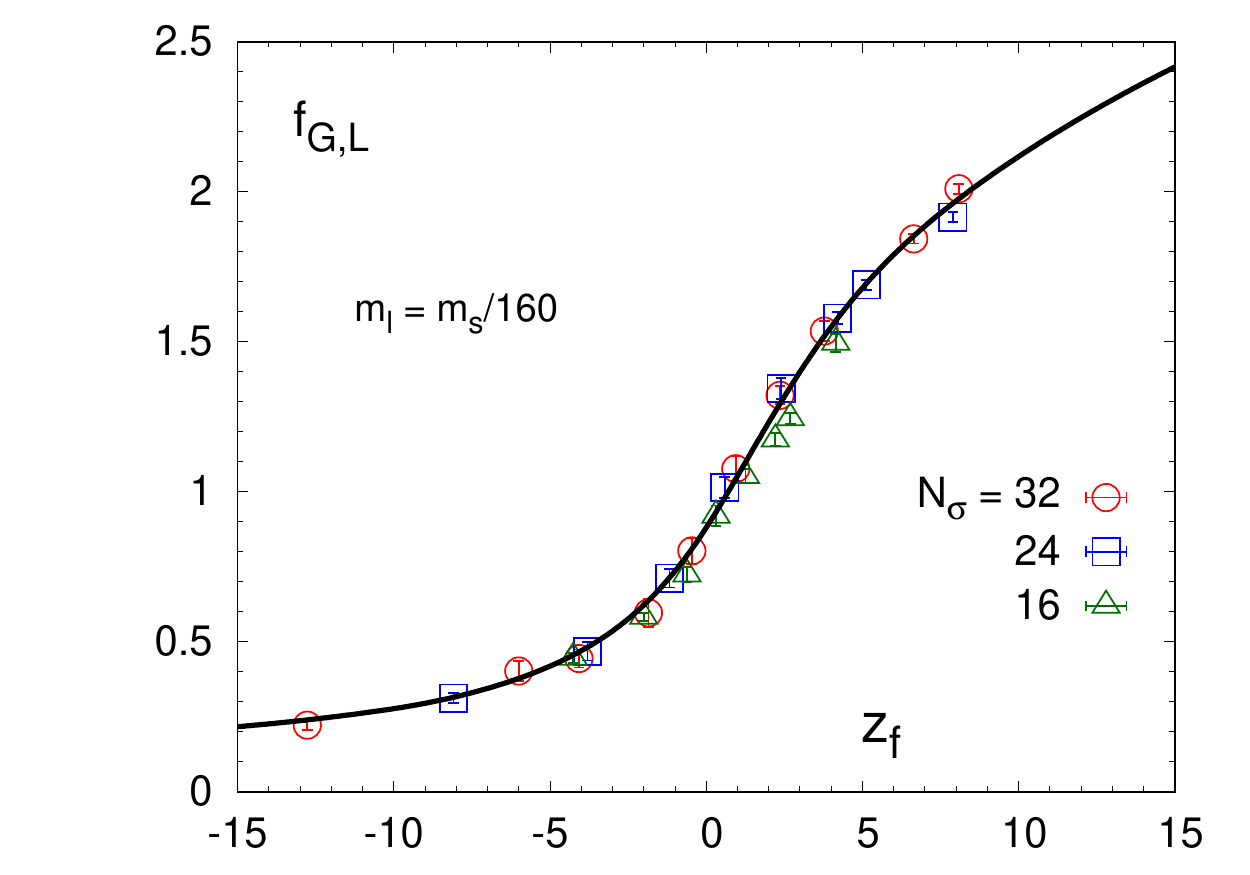}
\includegraphics[width=0.32\textwidth]{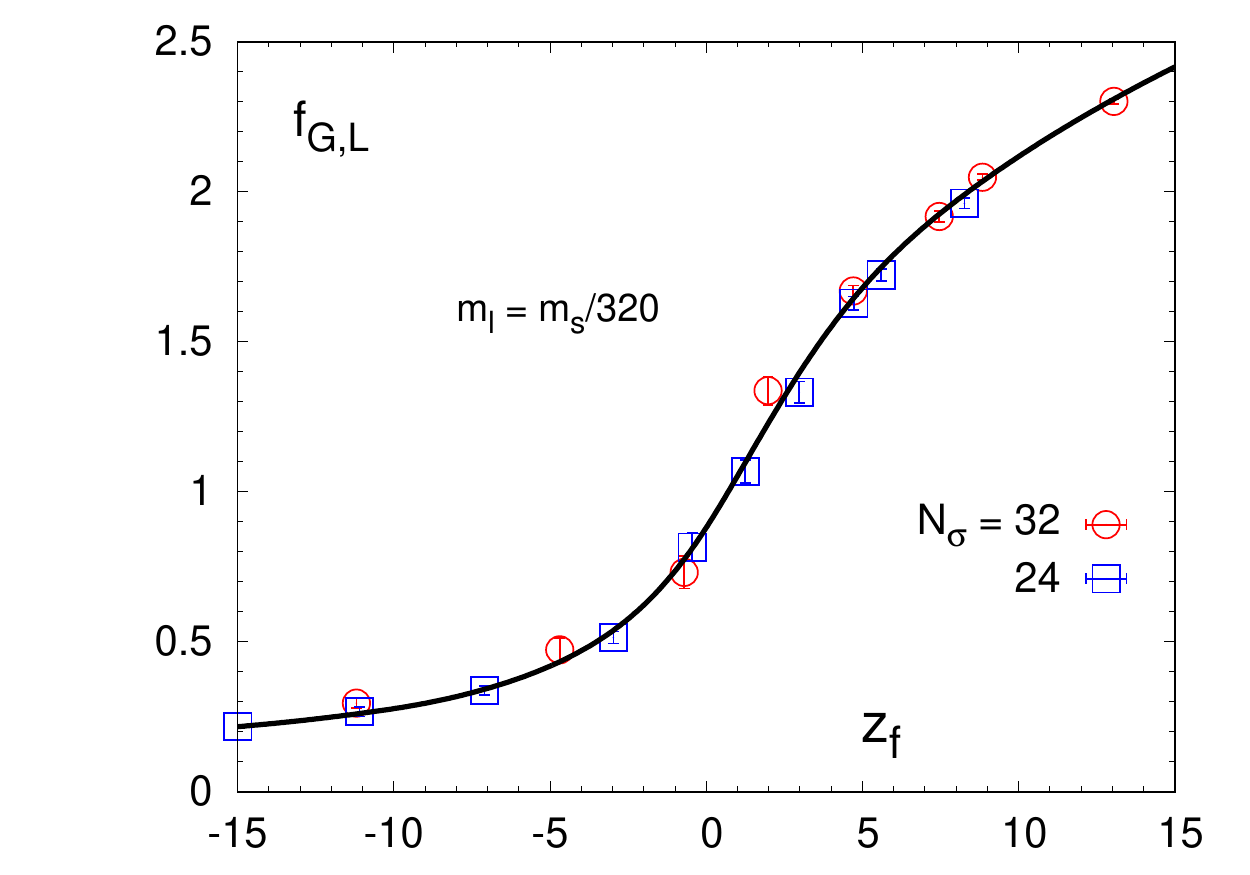}\\
\includegraphics[width=0.32\textwidth]{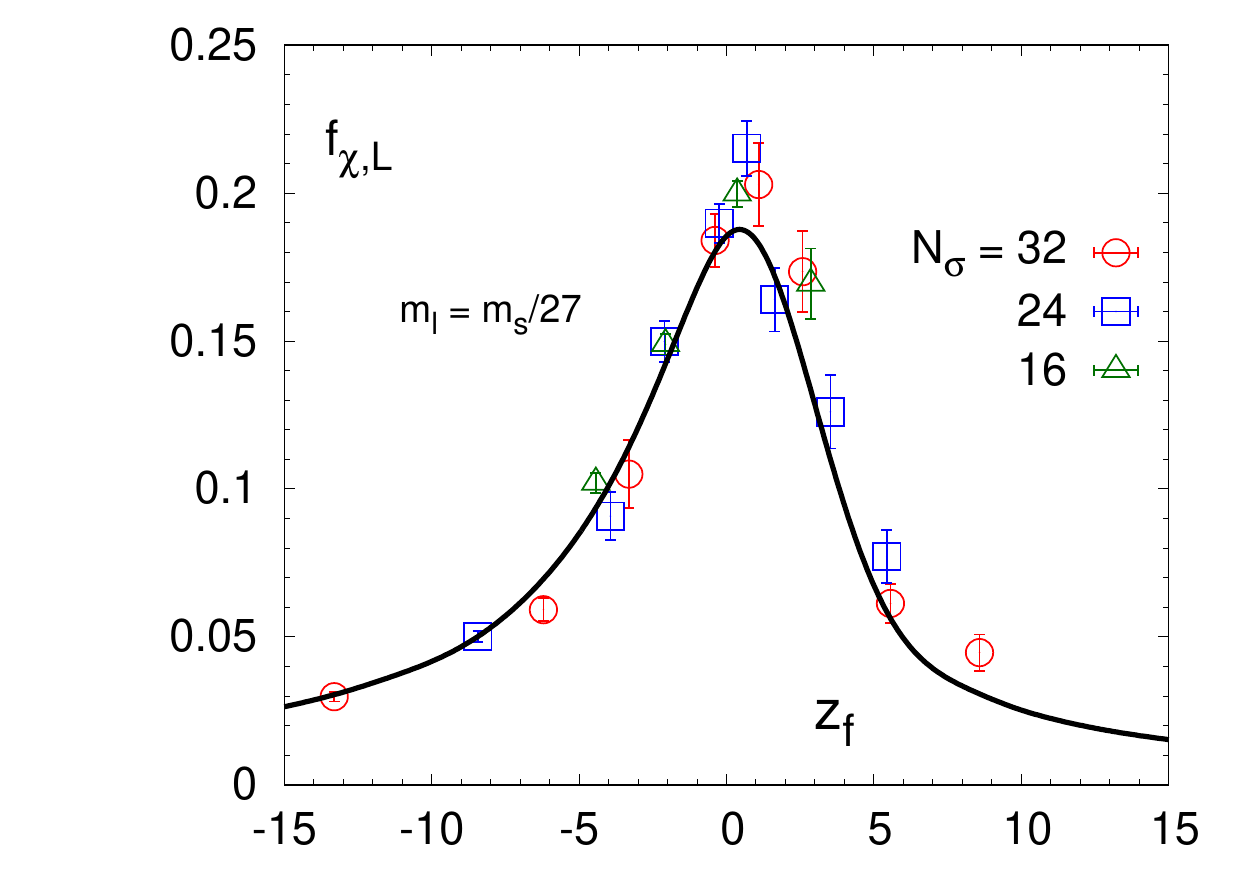}
\includegraphics[width=0.32\textwidth]{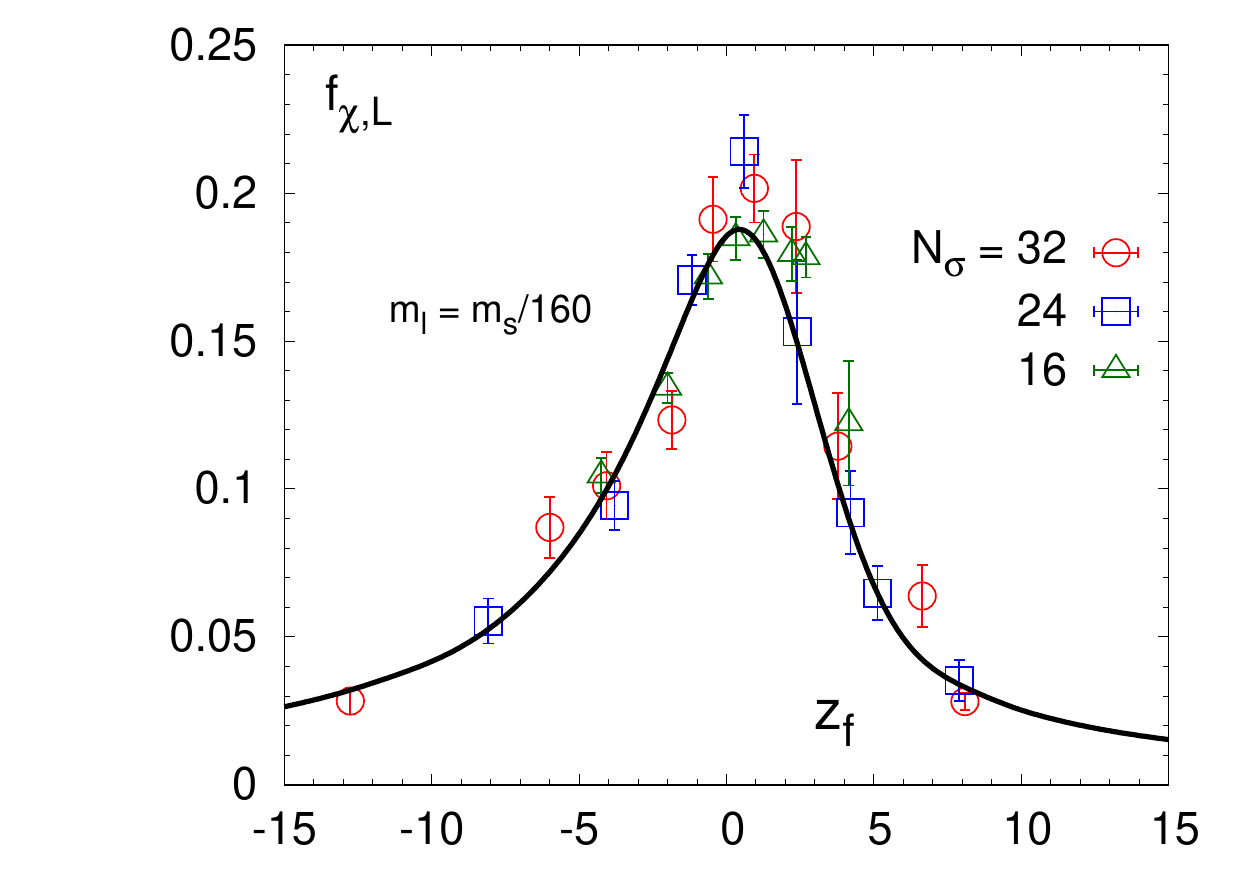}
\includegraphics[width=0.32\textwidth]{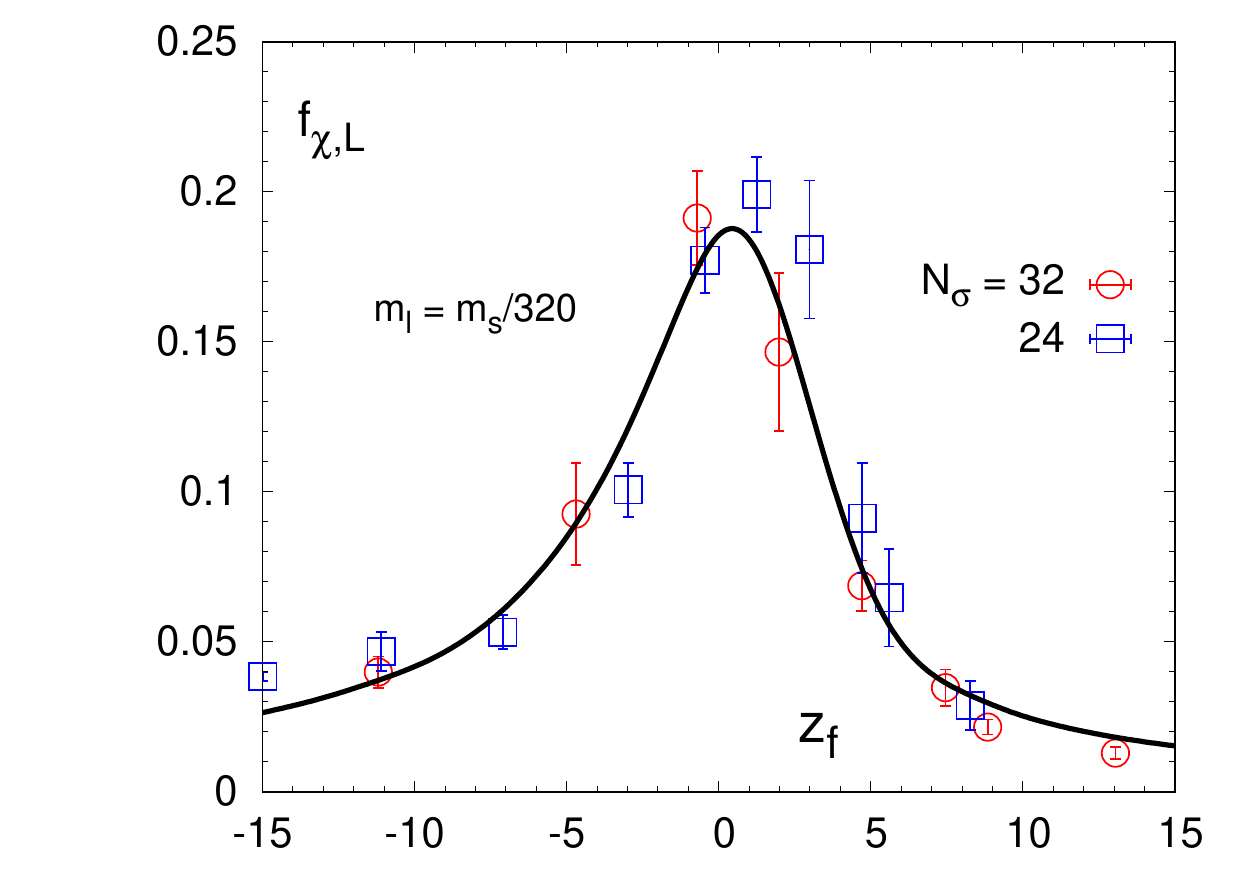}
\caption{Plots of finite size scaling functions for $M$ (upper row)
and $\chi_M$ (lower row). From left to right, columns give results for
light quark masses $m_s/27$, $m_s/160$, and $m_s/320$.
We subtract from $\chi_M$ the regular contribution, then data for both
$M$ and $\chi_M$ are re-scaled according to eq.~\eqref{eq:fitansatz} to isolate the
scaling functions. Black, solid lines are the universal scaling functions obtained 
from 3-$d$ improved Ising model calculations.}
\label{fig:fss}
\end{figure}

\begin{table}
\begin{tabularx}{\linewidth}{LCcCCCR} \hline\hline
$m_l$    & $A$        & $\TRW$ [MeV] & $z_0$      & $a_0$     & $a_1$    & $\chidof$ \\
\hline
$m_s/27$  & 0.1047(12) & 201.05(93)     & -1.126(33) & 0.072(22) & 2.00(31) & 2.46\\
$m_s/160$ & 0.0947(13) & 195.80(11)     & -1.073(35) & 0.107(25) & 2.11(44) & 1.60\\
$m_s/320$ & 0.0928(16) & 194.97(17)     & -1.026(30) & 0.145(34) & 2.12(49) & 1.70\\
\hline\hline
\end{tabularx}
\caption{Summary of fit parameter results for the joint fit given by
         eq.~\eqref{eq:fitansatz}. Temperature ranges for the fits in MeV
         are [191, 213], [186, 206], and [181, 202] for $m_s/27$, $m_s/160$,
         and $m_s/320$, respectively.} 
\label{tab:fitresults}
\end{table}

For the critical exponents $\alpha$, $\beta$, $\gamma$, and $\nu$ 
we use~\cite{Zinn-Justin:1999opn}
\begin{equation}
\alpha=0.1088,~~~~ \beta=0.3258,
~~~~\gamma=1.2396,
~~~~\text{and}~~~~
~~~~\nu=0.6304.
\end{equation}
Following the scaling behavior eq.~\eqref{eq:Mscaling} and
eq.~\eqref{eq:Bscaling},
we employ for $M$, $\chi_M$, and $B_4$ the ans\"atze
\begin{equation}\begin{aligned}\label{eq:fitansatz}
  M&= AN_\sigma^{-\beta/\nu}f_{G,L}(z_f)\; ,\\
  \chi_M&= A^2N_\sigma^{\gamma/\nu}f_{\chi,L}(z_f)+a_0+a_1t \; ,\\
  B_4&= f_B(z)\; .
\end{aligned}\end{equation}
We have included the leading regular corrections for $\chi_M$ that respect the
$\Z_2$ symmetry. This ansatz then corresponds to a five-parameter fit in
the non-universal parameters $A$, $\TRW$, and $z_0$ as well as the leading
regular coefficients $a_0$ and $a_1$. We perform first a joint
fit for the scaling functions $f_{G,L}$ and $f_{\chi,L}$. The results for $z_0$ and
$\TRW$ are then plugged into $f_B$, which serves as a consistency check.
All fits use $N_\sigma\geq24$ to reduce the effects of regular terms
and corrections-to-scaling, and we use a temperature range approximately
$\TRW\pm10$~MeV.

\begin{figure}
\centering
\includegraphics[width=0.45\textwidth]{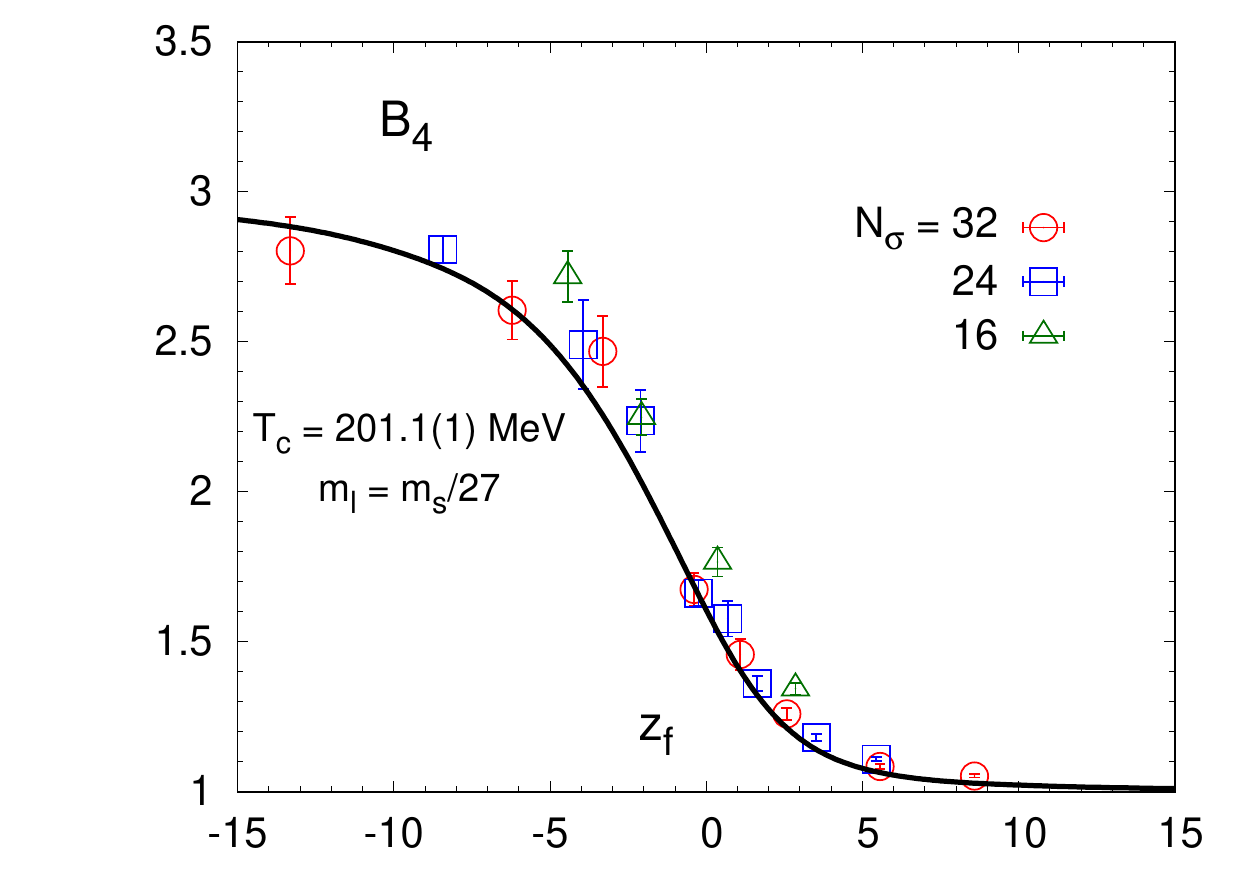}
\includegraphics[width=0.45\textwidth]{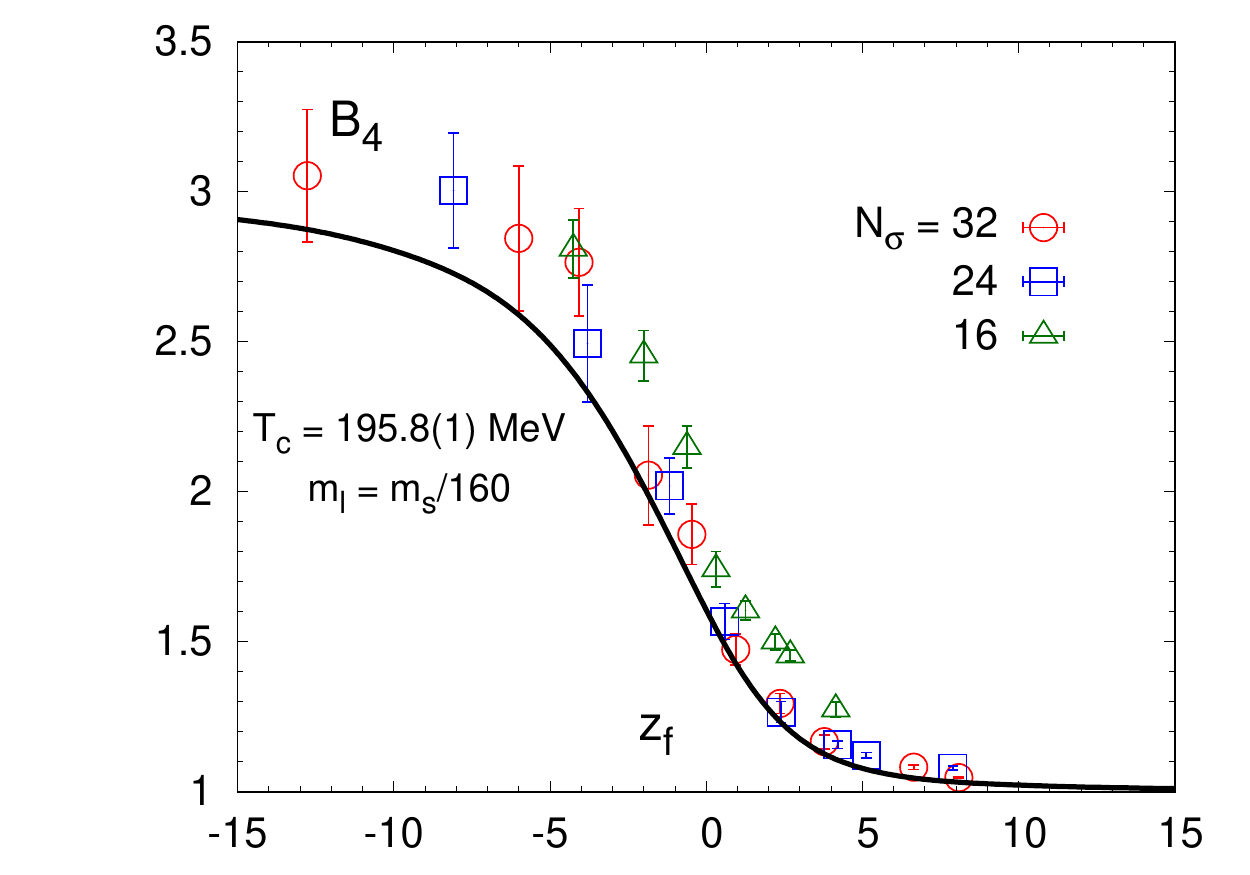}
\caption{Finite size scaling plots for $B_4$ for $m_s/27$ (left)
and $m_s/160$ (right).}
\label{fig:B4}
\end{figure}

In Fig.~\ref{fig:fss} we show the results of our finite size scaling fits
plotted against the finite size scaling variable $z_f$.
The top row shows fits for the $f_{G,L}$ and the bottom row shows fits for
$f_{\chi,L}$, and each column shows the result for the fits at different
quark masses.
We subtract the regular part from the $\chi_M$ data, then re-scale both
$M$ and $\chi_M$ to represent 
only the universal part as defined in 
eq.~\eqref{eq:fitansatz}. The universal functions are based on an improved 
3-$d$ Ising model calculation
and indicated by the solid, black curves. Results for
fit parameters are given in Table~\ref{tab:fitresults}. $\TRW$ exhibits
a statistically significant dependence on $m_l$, decreasing with
decreasing $m_l$. 
In Fig.~\ref{fig:fss} we see good agreement with the 3-$d$, $\Z_2$ expectation 
and no evidence of a first-order transition down to $m_l=m_s/320$. Fit results
for $B_4$ are shown in Fig.~\ref{fig:B4}. 
The fits fall near the data, serving as a consistency check.

% label is wrong in right plot. Does it come from same fit as before?
\begin{figure}
\centering
\includegraphics[width=0.45\textwidth]{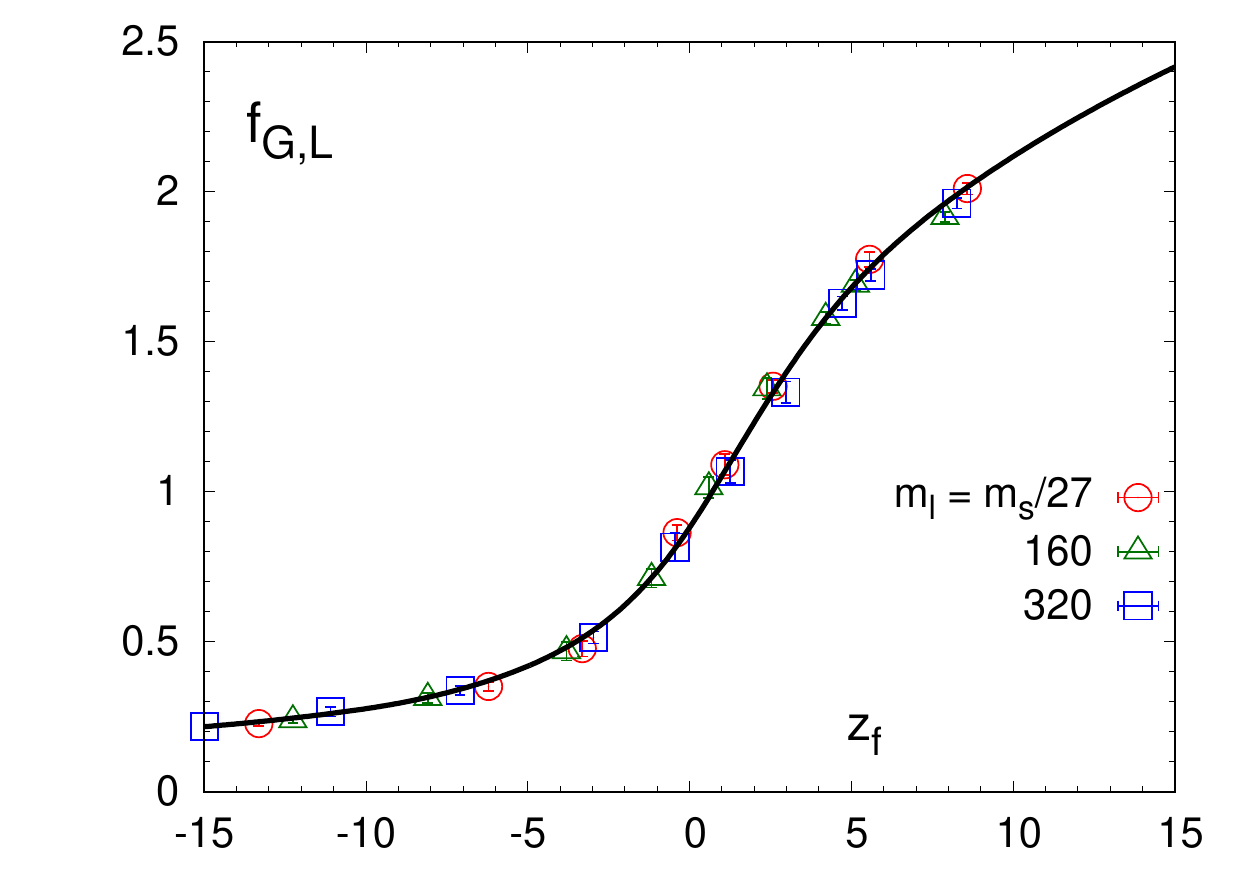}
\includegraphics[width=0.45\textwidth]{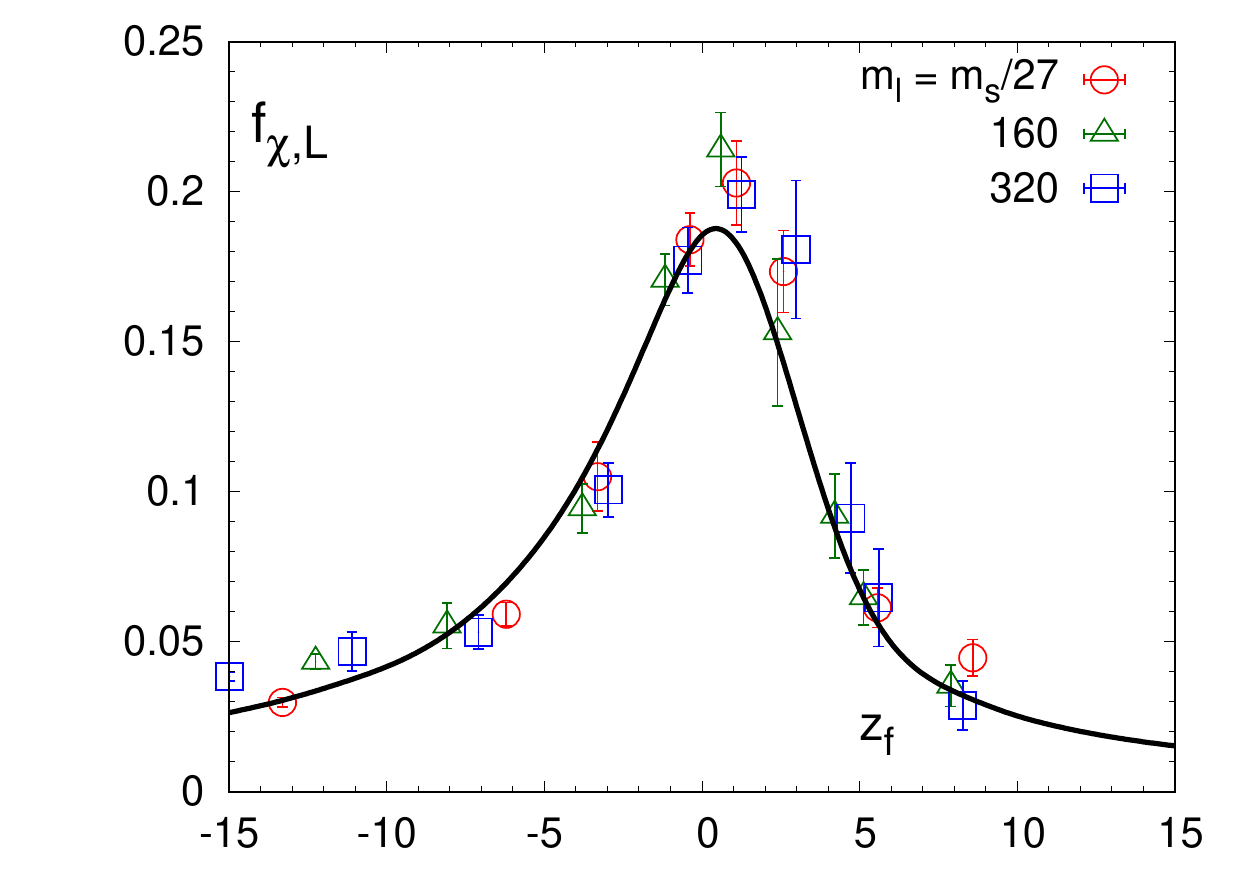}
\caption{Finite size scaling plots for re-scaled $M$ and $\chi_M$
at fixed $N_\sigma=24$ for different $m_l$. The universal scaling function is displayed as a
black curve.}
\label{fig:tscaleM}
\end{figure}

Results for the re-scaled $M$ and $\chi_M$ at fixed $N_\sigma=24$ down 
to $m_l=m_s/320$ are shown in Fig.~\ref{fig:tscaleM}. $\Z_2$
scaling fits are shown as black curves. As can be seen in these
plots, the data are consistent with the 3-$d$, $\Z_2$ scaling functions
down to $m_l=m_s/320$, showing no evidence for first-order behavior
even at our smallest pion mass at approximately 40~MeV. Our results are consistent with the findings of ref.\ \cite{Bonati:2018fvg}.

\begin{table}
        \begin{tabularx}{\linewidth}{LCR} \hline
                \hline
		$m_l/m_s$ & $T_{\rm infl.}^{\Delta_{ls}}$ [MeV] & $T^{\chidisc}_{\rm peak}$ [MeV] \\
                \hline
                1/27 &  202.6(7) & 202.6(2)  \\
                1/160 & 196.7(6) & 197.3(2)  \\
                1/320 & 195.9(7) & 194.6(6)  \\
                \hline\hline
       \end{tabularx}
	\caption{Chiral pseudocritical temperatures as determined by the location of
	         the inflection point in $\Delta_{ls}$ and peak in $\chidisc$ for the largest available volume $N_{\sigma}=32$.}
        \label{tab:z0Tc}
\end{table}

\begin{figure}
\centering
\includegraphics[width=0.48\textwidth]{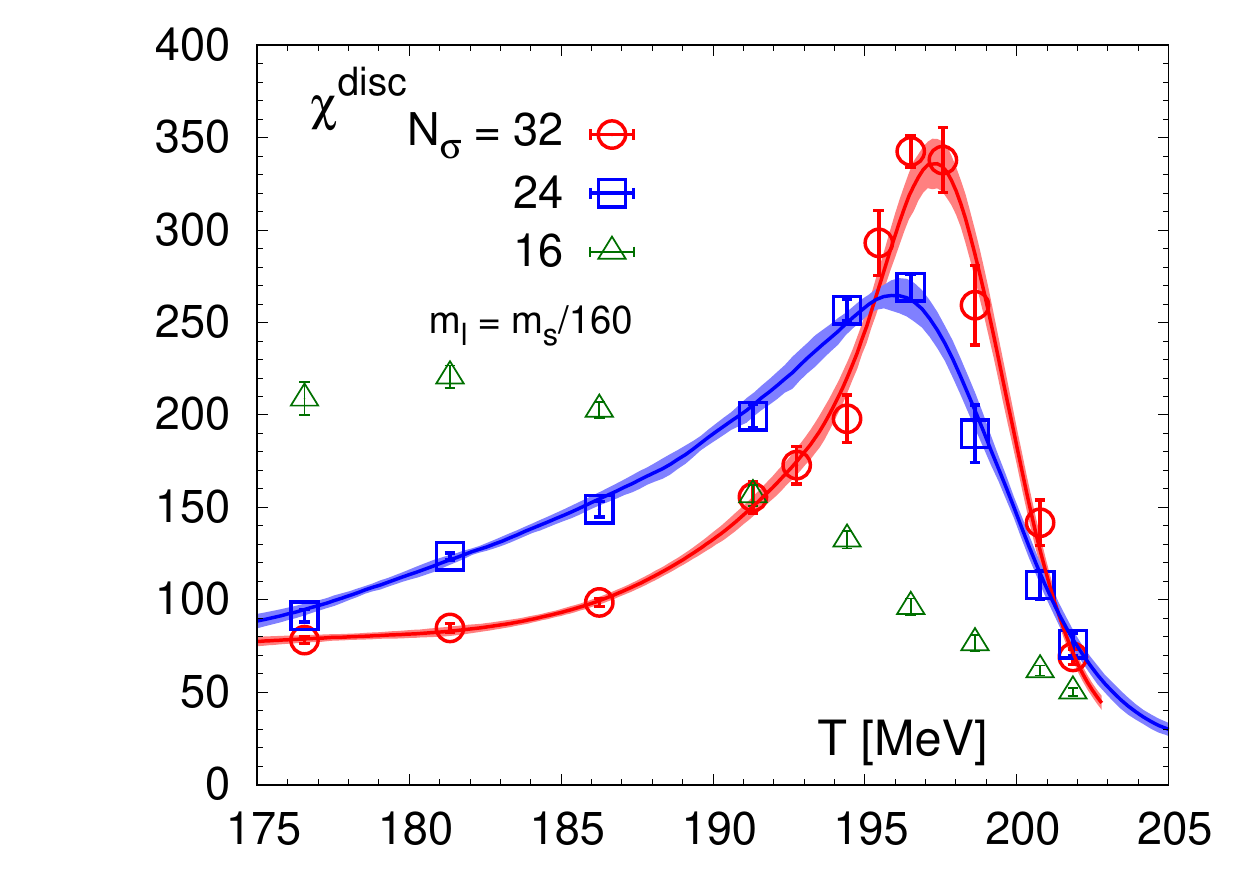}
\includegraphics[width=0.48\textwidth]{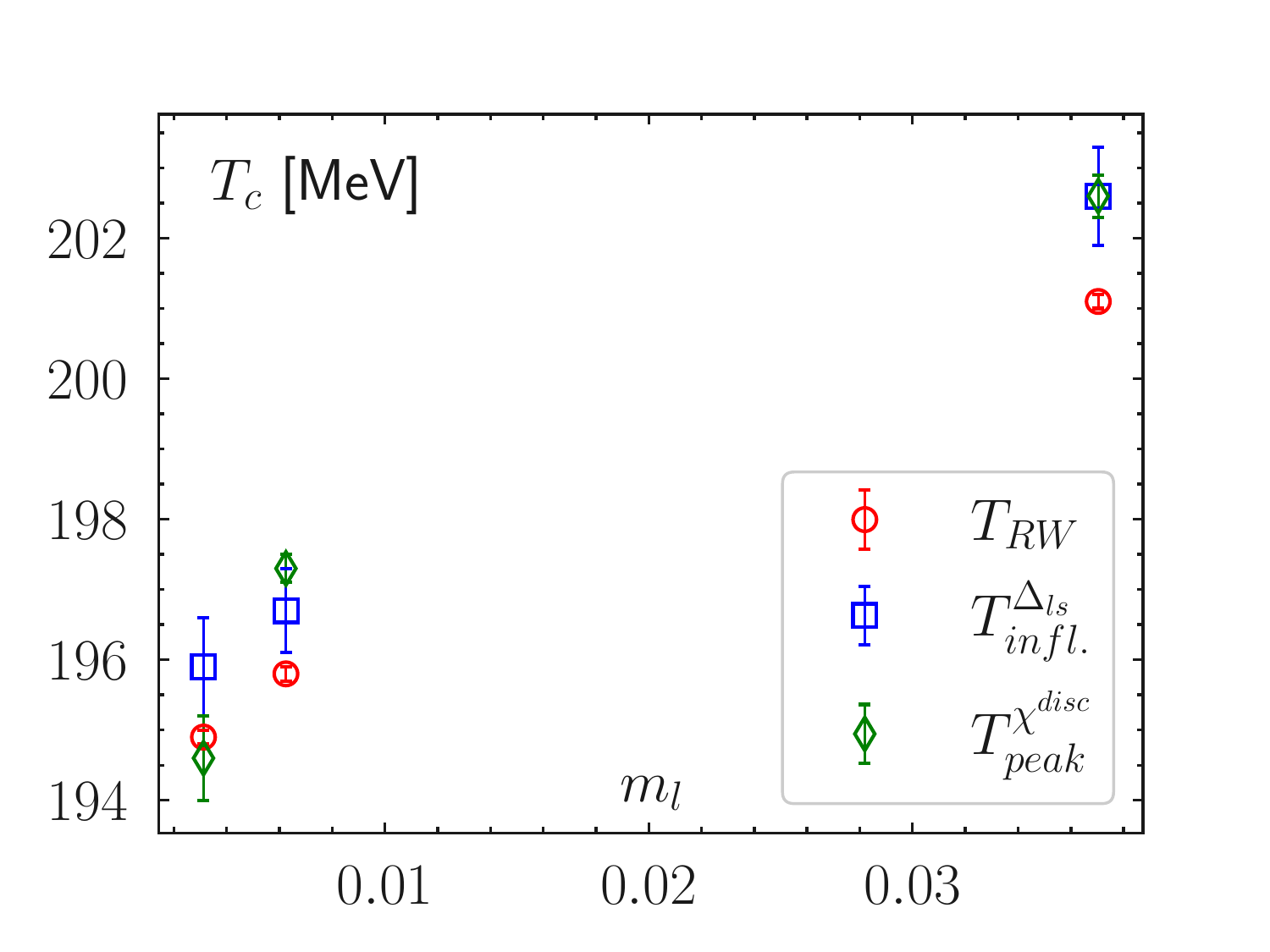}
\caption{Interplay between RW transition and chiral transition.
{\it Left:} Dependence of $\chidisc$ on $N_\sigma$ at $m_l=m_s/160$.
{\it Right:} $\TRW$ from Table~\ref{tab:fitresults} and $T_{pc}$ coming from the inflection point in $\Delta_{ls}$
and the peak in $\chidisc$ against $m_l$.} 
\label{fig:Z2O4}
\end{figure}

\subsection{Interplay between RW and chiral transition}
Finally we turn to the sensitivity of chiral observables to the RW transition.
In Fig.~\ref{fig:Z2O4} (left) we show $\chidisc$ at $m_l=m_s/160$. Instead of a
slight decrease of peak height with increasing $N_\sigma$, which is what
one would expect from $\O(N)$ models, there is instead an increase with
$N_\sigma$, which is what one expects from eq.~\eqref{eq:RGchidisc} in the vicinity of a critical point controlled by the 3-$d$, $\Z_2$ universality class. In
Fig.~\ref{fig:Z2O4} (right) we show the $\TRW$ obtained from fitting the $\Z_2$ scaling functions as listed in Table~\ref{tab:fitresults} along with pseudocritical
temperatures extracted using the $\Delta_{ls}$ inflection point and
the $\chidisc$ peak at our largest available volume $N_{\sigma}=32$ given in Table~\ref{tab:z0Tc}. There is a clear separation of temperatures at our largest
$m_l$, but at the smallest $m_l$ they are statistically compatible.
This may hint that the RW and chiral transitions coincide. In that
case, a larger symmetry group and universality class would be relevant.

\section{Summary and outlook}\label{sec:outlook}

The RW endpoint appears consistent with the 3-$d$, $\Z_2$ universality class down to
$m_\pi\approx40$~MeV, and calculations at imaginary $\mu$ set an upper bound
$\mcrit\leq40$~MeV also for a possible regime of first-order transitions in (2+1)-flavor QCD at vanishing chemical potential. The $\O(N)$ and $\Z_2$ transitions may coincide in the
chiral limit, resulting in a universality class different from both $\Z_2$ and
$\O(N)$.

\section*{Acknowledgements}

This work was supported in part through Contract No. DE-SC001270 with the U.S. 
Department of Energy, the Deutsche Forschungsgemeinschaft (DFG) through the CRC-TR 211
"Strong-interaction matter under extreme conditions", grant number 315477589-TRR 211, grant 05P18PBCA1 of the German Bundesministerium f\"ur Bildung und Forschung and the grant 283286 of the European Union.

\bibliographystyle{JHEP} % JHEP is the style requested by the PoS people
\bibliography{bibliography}

\end{document}